# Multimodal analogs to infer humanities visualization requirements

Richard Brath

Uncharted Software Inc.

**ABSTRACT**

Gaps and requirements for multi-modal interfaces for humanities can be explored by observing the configuration of real-world environments and the tasks of visitors within them compared to digital environments. Examples include stores, museums, galleries, and stages with tasks similar to visualization tasks such as overview, zoom and detail; multi-dimensional reduction; collaboration; and comparison; with real-world environments offering much richer interactions. Some of these capabilities exist with the technology and visualization research, but not routinely available in implementations.

**Index Terms**: K.6.1 [Management of Computing and Information Systems]: Project and People Management—Life Cycle; K.7.m [The Computing Profession]: Miscellaneous—Ethics

## 1 INTRODUCTION

Multimodal visualization for humanities subjects may seem to be a challenge due to the limitations of modalities – such as touch, smell and taste – however there are additional experiential aspects of cultural interpretation that are also lost in digitization, data transformation and computational representation. By observing some of these experiential aspects in the existing analog environments we can both confirm existing visualization interaction patterns and determine potential additional requirements for interactive representations of data and properties associated with multimodal cultural artifacts.

The contribution shows opportunities for extensions to visualization techniques, such as improved overview and clustering (Section 2); cues and collaboration from other visitors (3), and broader comparison tasks (4). These are discussed and situated to prior and future research (5).

## 2 SOME CONFIRMATIONS

Some real-world techniques match patterns promoted in visualization design.

### 2.1 Overview, Zoom, Details

The heavily cited Shneiderman visualization mantra is, "Overview first, zoom and filter, then details-on-demand," [1]. This pattern occurs in many real-world environments, such as the fashion retailer in Figure 1. which has a physical 3D arrangement organized to create a visual overview across the entire range of goods throughout the entire store from front to back [2]. Initial view and initial movement in the space provide overviewing [3] of the detail goods and reveals structure (e.g. gender is left and right, new/old are front/back, basics/specialties are side/center). Object positioning (orientation and location) and lighting (brightness) direct attention to aid tasks such as navigation, exploration and planning. Guidance may be provided by creating a hierarchy of pathways thereby directing movement (e.g. Ikea). Zoom and filter is the simple motion of walking to the proximity of a group of goods, with details on demand either by direct investigation of the artifact, or by conversation with the real-world agent (i.e. the staff). Note that the details are myriad, including close visual inspection, texture and form, meta-data (e.g. label indicating material, care, country of origin, price, and so on).

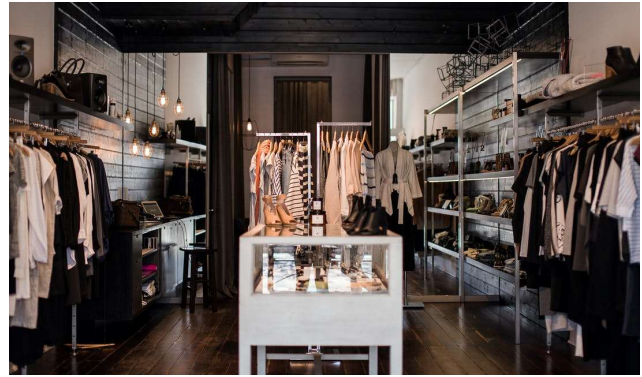

Figure 1: Fashion retail: Overview on entry, zoom by walking, details by inspection or verbal query.

Note that the digital equivalent of this experience (e.g. Amazon) provides interaction with the meta-data (filters) but no haptic interaction (What does the material feel-like? How will the material drape? Does it crease? Does it stretch?); nor derived attributes (How does this look in sunlight?). Furthermore, post-Covid evidence indicates people returning to real-world retail experiences, presumably indicating a desire for different experience in analog environments [4,5]. Arguably, current digital retail experiences provide a poor overview: consider the number of goods visible upon entry in Figure 1 vs. the online experience.

### 2.2 Multi-dimensional reduction

Clustering, and more broadly multi-dimensional reduction of high dimensional spaces into 2 or 3

dimensions, is a popular topic in visualization (e.g. [6,7,8]). It is also a commonly solved problem in real-world analogs. Figure 2, shows a series of bottles with a wide variety of characteristics, e.g. brand, alcohol content, source grain, frequency of use, etc., organized so that highly similar contents are closest together [9]. In this real-world analog, a human curator has organized the content to suit the needs (presumably facilitating fast visual search, quick retrieval of most popular items, and overviewing the breadth of beverages available).

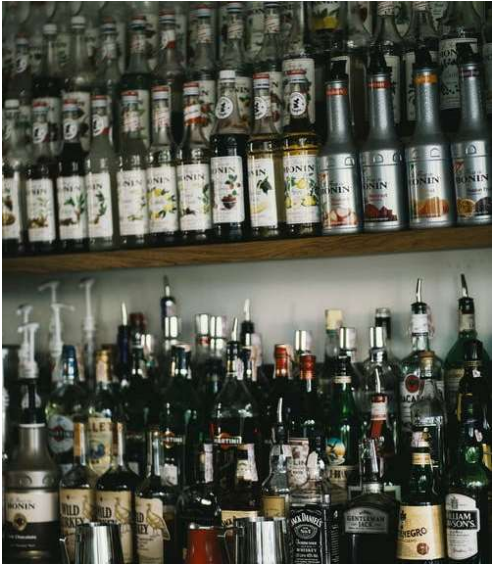

Figure 2: Bottles organized by muiltivariate similarity.

This clustering curation is central to organizing analog collections of artifacts such as museums and art galleries. Figure 3 shows a museum floorplan with rooms labeled across orthogonal topics such as *Korea*, *dinosaurs*, *toy soldiers*, *biodiversity* and *florals*, yet a visitor would have a high degree of confidence regarding the items they would find in each of these rooms.

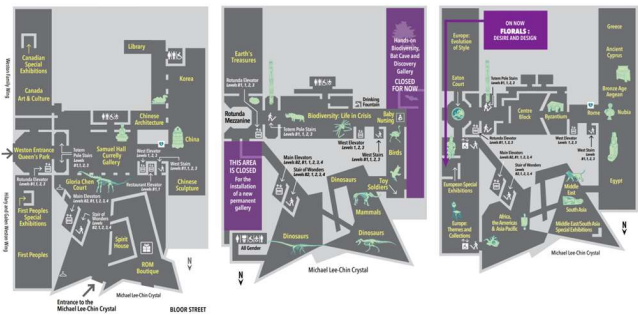

Figure 3: Musuem collection curation. Rooms are disparate, such as, Korea, Dinosaurs, Toy Soldiers, Biodiversity, and Florals.

While visual analytic systems may use unsupervised clustering, there are many unsolved problems, such as determining the number of relevant clusters to the task, generating appropriate labels, and creating interpretable visual layouts.

## 3 SOCIAL CUES AND COLLABORATION

In many real-world analogs there are strong collaborative interactions.

### 3.1 Serendipitous Eavesdropping or Collaborative Interpretation

Figure 4 shows an art gallery with many visitors [10]. Movement through the gallery requires recognition of other visitors, to avoid collision, and more generally provides additional information, such as which art works are more popular (number of people at a work), how to interact with a particular piece (are people appreciating it from afar or up close?), what content within an artwork is being engaged with (where is the visitor's attention with regards to a work?), what might people be saying about a particular work (how are individuals commenting about a particular piece?), and so on. As opposed to serendipity across cultural artifacts [11]; this is serendipity across viewers, visible to the information flaneur [12].

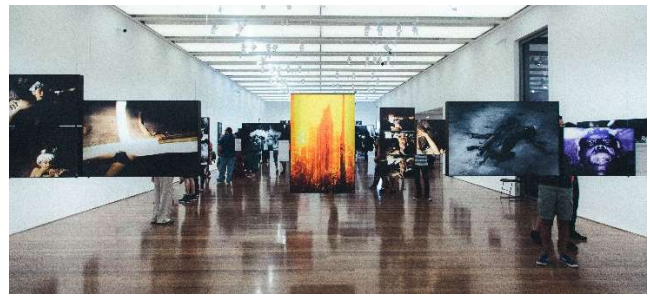

Figure 4: Crowded galleries for art appreciation and serendipitous interpretation.

Although galleries can be viewed with less people (e.g., early morning, or in the last few days of a show), the rich interaction between humans is important—private galleries are particularly busy at show openings: people want to be with other people in addition to the art, e.g. [13].

### 3.2 Collaborative Selection

Content engagement may be directed and encouraged by trusted relationships, such as friends or staff, as shown in Figure 5 [14,15].

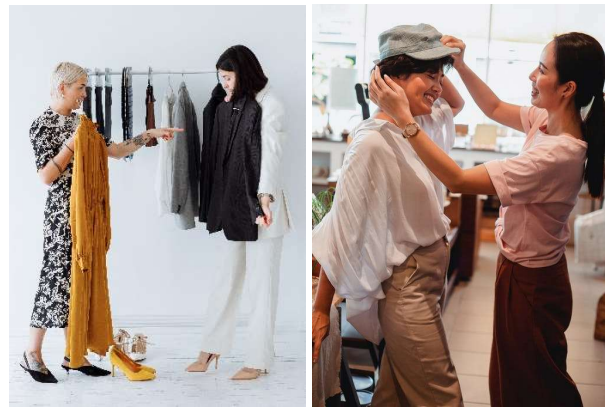

Figure 5: Collaborative selection.

The visitor may be hesitant but may be otherwise encouraged to make and commit to a selection with the additional input of a peer. This is a form of synchronous collaborative user guidance e.g. [16].

### 3.3 Collaborative Sorting

Content can be directly engaged with, and visitors may reorganize content, if permitted. Figure 6 shows a woman selecting and organizing fruit [17]. With each selection, various criteria are evaluated, such as size, color, texture, firmness, surface anomalies, shape, smell and so forth. This is not simple filtering—a confluence of metadata is evaluated, and a single piece of fruit could be accepted or rejected from a combination of criteria rather than thresholding on values. Over time, successive rejections by successive visitors will cause items to become sorted—one area in the bin will likely contain highly rejected fruit (e.g. rotten), whereas another portion of the bin may contain less desirable fruit (e.g. blemished). This fine-grain sorting is completely unavailable in digital counterparts.

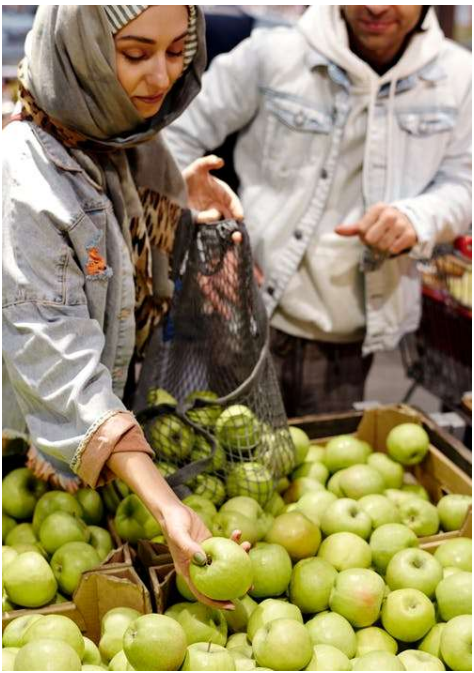

Figure 6: Collaborative sorting.

## 4 COMPARISON AND COORDINATION

Comparison is a frequent task in visualization, with subjects required to assess the relative sizes, positions, and so forth between marks within a visualization e.g., [18,19,20]. The richness, complexity, and nuances of cultural objects invites many comparisons and are thus frequently placed in proximity to facilitate comparison – although not necessarily comparison of quantitative data.

### 4.1 Curated comparison

A museum curator or shopping display may deliberately place items adjacent to each other to direct comparison, such as paired artworks in Figure 7 [21].

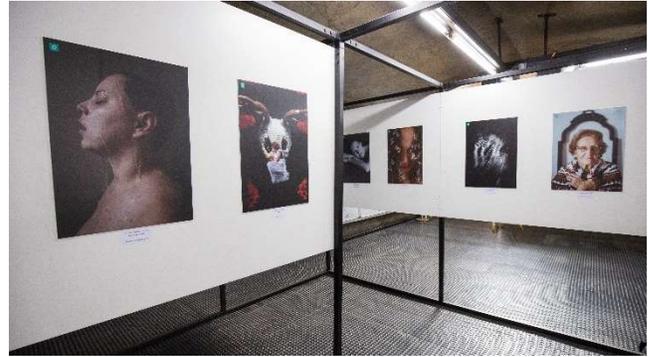

Figure 7: The curator has paired art works facilitating comparison.

In Figure 8, the curator has deliberately placed mid-century indigenous art by Karoo Ashevak (foreground sculpture) juxtaposed with contemporary mainstream art (paintings on wall) [22]. This uncommon pairing by the curator forces criticality onto the viewer to consider indigenous art in the context of its broader culture, not merely in comparison to other indigenous art.

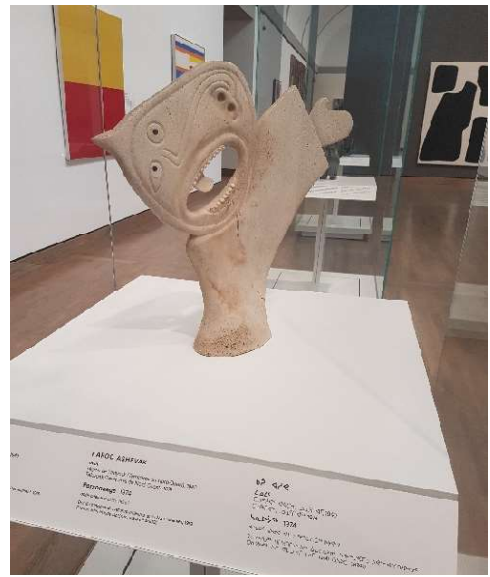

Figure 8: Sculptures by indigenous artists curated together with contemporary mainstream art forcing the viewer to compare the relation between the two.

Symmetry and repetition invite comparison, such as choreographed routines, as in Figure 9 [23]. While choreographed attributes could be directly visualized and compared, e.g., such as positions and angles and deviations among performers; the viewer may also attend to other comparisons, such as performers' expression, lighting, hairstyle, complexion, and so forth.

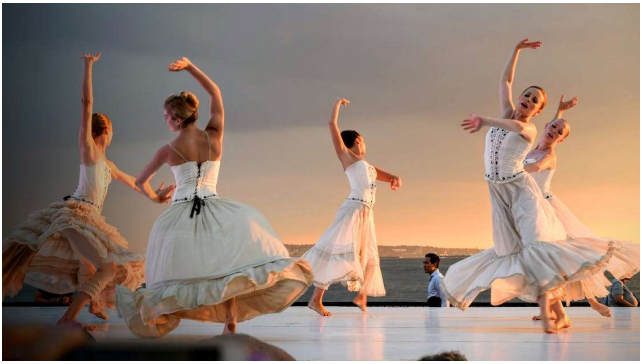

Figure 9: Comparison could be between choreographed attributes such as pose and body angles, or features such as expression, lighting, hair style, and so on.

### 4.2 Ad hoc comparison

The participant, however, may wish to do their own comparisons. In real-world analogues, these comparisons can be done easily in situations, such as retail stores, by selecting and dragging objects of interest into proximity to each other, such as Figure 10 left [24]. In addition to side-by-side comparison, the person can view objects from many angles, lighting conditions, and so on.

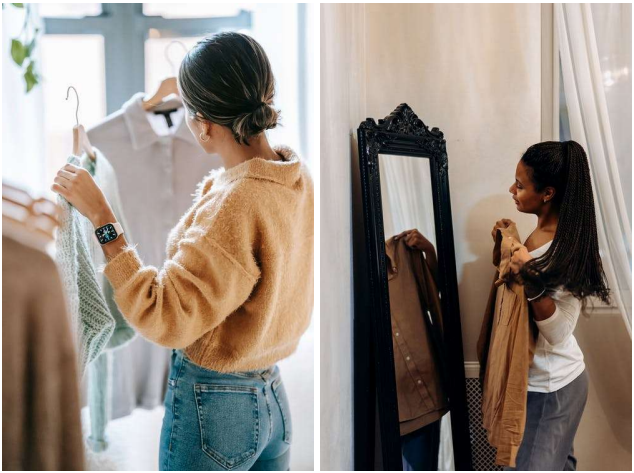

Figure 10: Object comparisons, e.g., against each other in bright light; in relation to subject.

This comparison can be between subject and object, such as viewing the object superimposed on the subject in a mirror to assess how the object appears in relation to the subject's skin tone, hair color, body shape and so on (Figure 10 right [25]). Dudley [26] notes the importance of direct experience with the materiality of museum artifacts, e.g. understanding the artifact in relation to self.

### 4.3 Coordination and Composition

The task may be the assembly of dissimilar objects into a composition. This may be challenging in digital environments and visualizations as the metadata may be quite different, such as jackets, shirts, ties and belts in Figure 11 [27]. The real-world analog can literally be dragging objects overtop each other.

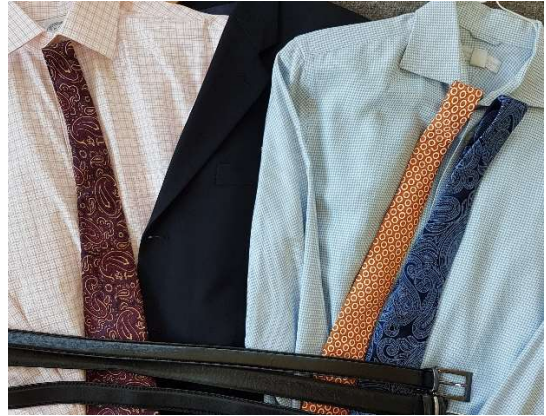

Figure 11: The non-curator has assembled to disparate objects to assess potential compositions.

## 5 DISCUSSION, PRIOR AND FUTURE WORK

Aspects such as serendipity, generosity, criticality and guidance have been discussed in [28]. Generous interfaces by Whitelaw [29] describes rich interactive interfaces with many similarities to the examples above, such as "show everything"; multivariate overviews, samples and context, and high-quality primary content.

***Multi-media and labels***. Prior work in multimedia indicates improved task performance when using multiple channels. For example, learning and problem-solving tasks are improved when textual instructions are directly integrated into diagrams [30]. Other studies in multimedia have shown improved performance when information is provided through both text and images with close spatial positioning and/or visual linkages between the two [31].

At a minimum, prior multi-media work indicates the importance of textual labels which can be utilized at the level of individual items or across clusters. Whitelaw's examples use simple term frequency to extract single word labels for clusters, but this is an ongoing technical challenge. Techniques such as Latent Dirichlet Allocation (LDA) return ranked word lists not singular descriptive words. Summarization techniques may be challenged to reduce to one-word labels such as Figure 3. Even reduced to singular words, challenges remain for integrating labels directly into the visualization while maintaining the legibility of labels and not obscuring the data visualization marks.

***Game-Quality Rendering***. Visualizations typically transform data into a small palette of visual attributes, such as size, color, brightness and position – e.g., libraries such as D3.js, and end-user tools such as PowerBI or Tableau. Advanced properties such as lighting, shading and materials are not used.

However, attributes of cultural artifacts, such as material or motion properties important in their analog interactions and comparisons (e.g., Figure 9,10). These could be retained and depicted realistically within modern high-performance game engines, even if abstracted from their original objects into swatches and visualization layouts. For example, the myriad of cloth properties associated with historic clothing could directly represented as a virtual physical visualization using an underlying game engine. Some VR/AR visualizations use game engines, and as advanced effects are built-in to these engines, features such as flames, particle systems, glow and shadows are used, e.g. [32,33]; and some humanities visualizations create more sensuous effects such as animated sorting of piles of coin images [34].

Gaming engines with high quality 3D, motion, material properties, lighting effects, and so forth might be dismissed as unnecessary 3D [35] and distracting for low dimensional data; yet, the highly multivariate textural and motion qualities associated dance, music, costumes and other objects could be directly depicted, interpreted and compared.

***Augmented Reality***: Willet et al [36] outline superpowers as a source of inspiration for interactive visualization techniques. Some of these interactions are highly relevant to digital humanities visualizations. For example, humanities artifacts may not be directly accessible for close (touch) inspection, but could be investigated using AR x-ray vision.

Or, *RealitySketch* [37], an AR system which generates predictive overlays, such as paths associated with physics (e.g. a pendulum), could be extended to other physical and temporal oriented humanities applications, e.g. dance prediction, to show expected and un-expected movements.

***Communication:*** Engaging with cultural artifacts is highly enriched in the presence of other individuals, changing our collective understanding and response to these artifacts. Understanding other individuals' engagement with artifacts becomes important, ranging from visible voyeur to direct simultaneous engagement with the individuals and the artifacts.

There are many existing systems that provide some capabilities, e.g. Disqus, a comment service for web pages, allowing users to read and post commentary; or Discord, allowing real-time instant messaging service enabling discussions across a topic; or Unity, a game engine with a service for real-time cross-platform communication. In all cases, the artifact being discussed is separate from the conversation requiring descriptions to cross reference. Direct collaborative markup of the visualization allows more directed conversations to particular aspects of the content, e.g. [38,39].

***Comparison and Workspaces:*** Comparison is a frequent task in visualization, *but* is typically highly constrained to objects within a singular visualization components. With many visualization systems made of multiple panels and coordinated interactions, the notion of dragging an object in one panel to another panel is impossible. In a museum or gallery, the curator has similarly imposed non-comparison, but the retail shopping example indicate visitors require agency to assemble and compare disparate artifacts. Furthermore, the visitor may wish to compare aspects of items across stores, not simply solved in digital environments. Some workspaces allow for content to move between panes (e.g. cut and paste), although it is non-obvious how this might work in visualization systems and this seems to be an area for future research. There are opportunities for visual interfaces to create unexpected pairings to induce correlations for the observers [40], helping to create a critical context.

## 6 CONCLUSION AND NEXT STEPS

The purpose of this paper is to raise a breadth of issues in the gap between real-world analog environments and digital analytical visualization environments; largely through observation of tasks in real-world tasks and goals as a means for prompting potential interactions and requirements. Some of these issues are already solved in digital environments and prior research, although not necessarily used in routine visualization. To bring these capabilities routinely into visualization, is perhaps more of an issue of ease of implementation (visualization toolsets do not have these features); as well as cultural expectations – e.g. art websites don't necessarily have an expectation to reveal visitors to other visitors particularly in a period of increasing privacy legislation.